\documentstyle[aps,preprint,epsfig,floats]{revtex}
\tighten

\newcommand{\be}{\begin{equation}}
\newcommand{\ee}{\end{equation}}
\newcommand{\bea}{\begin{eqnarray}}
\newcommand{\eea}{\end{eqnarray}}
\begin{document}
\draft


\vspace{4cm}

\title{
Production of light pseudoscalars \\
in external electromagnetic
fields by the Schwinger mechanism}

\author{
J. A. Grifols $^a$ , Eduard Mass{\'o} $^a$
and Subhendra Mohanty $^b$}

\address{
$^a$ Grup de F{\'\i}sica Te{\`o}rica and IFAE,
Universitat Aut{\`o}noma de Barcelona, \\
08193 Bellaterra, Barcelona, Spain.}

\address{
$^b$ Physical Research Laboratory, \\
Navrangpura, Ahmedabad - 380 009, India.}

\maketitle
\vspace{2cm}
\begin{abstract}
We calculate the
probability of the decay of external inhomogeneous electromagnetic fields
to neutral pseudoscalar particles that have a coupling to two photons.
We also point out that our estimate for axion emission in a previous
paper was incorrect.
\end{abstract}

\pacs{14.80.Mz, 11.30.Er,12.20.Ds,13.40.-f}


\section{Introduction}

The Schwinger mechanism is a non-perturbative process by which an infinite
number of zero frequency photons can decay into electron-positron pairs
\cite{schwinger}.
In this paper we show that this mechanism can be generalized to study
the production of other kinds of
light particles from intense electromagnetic (EM) fields. The light
particle that we consider is a pseudoscalar (PS) having a coupling to
two photons.

In Section II we derive the formula for the decay
of classical background fields
into  PS particles. This is achieved by integrating out the
particle fields from the total Lagrangian to obtain the effective
action of the classical background fields. The imaginary part of the
effective Lagrangian is related to the probability of
decay of classical background fields into
particles. In Section III, we derive, from the usual coupling of the
PS to two photons, the specific interaction Lagrangian
that should be used in the general formalism of Section II in order
to account for vacuum decay into PS. For static EM fields, we show
that a necessary condition is that the fields are inhomogeneous.
In Sections IV, V, and VI, we explicitly calculate the PS production
in a variety of situations.
Specifically we consider a dipole magnetic field, a cylindrical 
capacitor, and a spherical
capacitor.
A final section is devoted to the conclusions.


\section{Decay of classical background fields into particles}

We start with the action for the pseudoscalar $\phi$ (mass $m$) coupled
to the background ${\bf E}$ and ${\bf B}$ fields of the general
form
\bea
S[ \phi, {\bf E}, {\bf B}]
&=& \int d^4 x~
        \frac{1}{2} \phi(x)
        \left[ - \partial^2 - m^2 + f(x)
        \right] \phi(x)
\label{eqn1}
\eea
where $f(x)$ is some scalar function of ${\bf E}$ and ${\bf B}$
fields.
From (\ref{eqn1}) we obtain the effective action for the
background ${\bf E}$ and ${\bf B}$ fields formally as
\bea
e^{i S_{eff}[{\bf E}, {\bf B}]}
&=& \int {\cal D} \phi~
        e^{i S[\phi, {\bf E}, {\bf B}]}
\label{eqn2}
\eea
The effective Lagrangian for the ${\bf E}$ and ${\bf B}$ fields
can be related
to the Green's function of $\phi$ in external ${\bf E}$ and ${\bf B}$ fields
as follows.
Differentiate (\ref{eqn2}) by $m^2$
\bea
i \frac{\partial S_{eff}[{\bf E}, {\bf B}]}
        {\partial m^2}
&=&
- \frac{ \int {\cal D} \phi~ \phi^2~
        e^{i S[\phi, {\bf E}, {\bf B}]} }
        { \int {\cal D} \phi~
        e^{i S[\phi, {\bf E}, {\bf B}]} }
\nonumber \\
&=& - \frac{1}{2} \int d^4 x~
        G \left( x, x; {\bf E}, {\bf B} \right)
\nonumber \\
&=& - \frac{1}{2} \int d^4 x~
        \int \frac{d^4 p}{(2 \pi)^4}~
        G \left( p; {\bf E}, {\bf B} \right)
\eea
The effective Lagrangian of the background fields is therefore
formally given by the expression
\bea
{\cal L}_{eff} [{\bf E}, {\bf B}]
&=& \frac{i}{2}
        \int dm^2 \int \frac{d^4 p}{(2 \pi)^4}~
        G \left( p; {\bf E}, {\bf B} \right)
\label{eqn4}
\eea
The probability of external ${\bf E}$ and ${\bf B}$ fields to decay
into quanta of $\phi$ is related to the imaginary part
of ${\cal L}_{eff}$ as follows
\bea
P &=& 1 - \left< 0 \left|~
        e^{i S_{eff}[{\bf E}, {\bf B}]}
        ~\right| 0 \right>
\nonumber \\
&=& 1 - \exp
        \left[- \,
        2~\mbox{Im} \int d^3 x~ dt~
        {\cal L}_{eff}[{\bf E}, {\bf B}]
        \right]
\label{eqn1.5}
\eea
In the case that this probability is small, we can
write the probability density $w$ (per unit volume and unit time)
approximately as
\be
w=  2~\mbox{Im}  {\cal L}_{eff}[{\bf E}, {\bf B}]
\label{w}
\ee

We now give the general procedure for obtaining the effective action
of the background fields by calculating the Green's function
of $\phi$ in background ${\bf E}$ and ${\bf B}$ fields following
the method of Duff and Brown\cite{duff}.

The effective Lagrangian can be calculated by this method
if the background fields contained in $f(x)$ in the interaction Lagrangian
\be
{\cal L}_I (x) = {1\over 2} f(x) \phi^2(x)
\label{f}
\ee
can be expanded in a
Taylor series near some reference point $\bar{x}$.
Expanding $f(x)$ near $x = \bar{x}$,
\bea
f(x)
= \alpha (\bar{x})
        + \beta_{\mu} (\bar{x}) (x - \bar{x})^{\mu}
        + \gamma^2_{\mu \nu} (\bar{x}) (x - \bar{x})^{\mu} (x - \bar{x})^{\nu}
+ ...
\label{eqn1.6}
\eea
\bea
\alpha (\bar{x}) &=& f(\bar{x}),
\hspace{10mm}
\beta_{\mu} (\bar{x})
= \left( \frac{\partial f}{\partial x^{\mu}} \right)_{x = \bar{x}},
\hspace{10mm}
\gamma^2_{\mu \nu} (\bar{x}) = \frac{1}{2}
        \left( \frac{\partial}{\partial x^\mu}
        \frac{\partial f}{\partial x^\nu} \right)_{x = \bar{x}}
\nonumber
\eea
The equation for the Green's function for the $\phi$  field
is given by
\bea
\left[
        \partial_x^2 + m^2 - \alpha
        - \beta_{\mu} (x - \bar{x})^{\mu}
        - \gamma^2_{\mu \nu} (x - \bar{x})^{\mu} (x - \bar{x})^{\nu}
\right]
G(x, \bar{x})
&=& \delta^4 (x - \bar{x})
\eea
In momentum space
\be
(x - \bar{x})^{\mu}
        \rightarrow
        - i \frac{\partial}{\partial p_{\mu}}
\ee
and the equation for the Green's function in momentum space is
\bea
\left[
- p^2 + m^2 - \alpha
        + i \beta_{\mu} \frac{\partial}{\partial p_{\mu}}
        + \gamma^2_{\mu \nu}
                \frac{\partial}{\partial p_{\mu}}
                \frac{\partial}{\partial p_{\nu}}
\right]
G(p) &=& 1
\label{eqn8}
\eea
We choose as an ansatz for the solution $G(p)$  the form
\bea
G(p)
&=& i \int_0^{\infty} ds~
        e^{- i s (m^2 - i \epsilon)}
        e^{i p_{\mu} A^{\mu \nu} p_{\nu} + B^{\mu} p_{\mu} + C}
\label{eqn9}
\eea
where $A(s), B(s) $ and $C(s)$ are to be determined. They must satisfy
the boundary condition
in the case of vanishing external fields,
i.e. when $\alpha, \beta, \gamma \rightarrow 0$
\be
A^{\mu \nu} \rightarrow s~ g^{\mu \nu},
~~~~~~
B^{\mu} \rightarrow 0,
~~~~~~
C \rightarrow 0
\label{bc}
\ee
and in this limit we should obtain
\be
G(p)
= i \int_0^{\infty} ds~
        e^{- i s m^2 + i s p^2}
= \frac{1}{m^2 - p^2}
\ee
i.e., the free particle Green's function.

To solve for $A,B$, and $C$ we
insert ansatz (\ref{eqn9}) in (\ref{eqn8}). We have
\bea
i \int_0^{\infty} ds~
        \Big[
          &-& \mbox{\boldmath p}^2 + m^2 - \alpha
          + i \mbox{\boldmath $\beta$} \cdot
                \left( 2 i \mbox{\boldmath A} \cdot
                \mbox{\boldmath p} + \mbox{\boldmath B} \right)
               + \left( 2 i
                        \mbox{\boldmath p} \cdot
                        \mbox{\boldmath A}
                        + \mbox{\boldmath B}
                   \right) \cdot
                {\mbox{\boldmath $\gamma$}}^2 \cdot
                   \left( 2 i
                        \mbox{\boldmath A} \cdot
                        \mbox{\boldmath p}
                        + \mbox{\boldmath B}
                   \right)
\nonumber \\
\hspace{2cm}
               &+& 2 i~ \mbox{tr}
                (
                {\mbox{\boldmath $\gamma$}}^2 \cdot
                \mbox{\boldmath A}
                )
        \Big]
        \exp \left\{ - i s m^2 + i \mbox{\boldmath p} \cdot
                        \mbox{\boldmath A} \cdot
                        \mbox{\boldmath p}
                        + \mbox{\boldmath B} \cdot
                        \mbox{\boldmath p}
                        + \mbox{C}
                \right\}
              = 1
\label{eqn10}
\eea
Equation (\ref{eqn10}) has the general form
\bea
\int_0^{\infty} ds~ g(s)~ e^{- h(s)} &=& 1
\eea
whose solution is
\bea
g(s) &=& \frac{\partial h(s)}{\partial s}
\label{eqn12}
\eea
with $h(0) = 0$ and $h(\infty) = \infty$.
Using the form of the solution (\ref{eqn12}) for equation (\ref{eqn10})
\bea
        i \Big[
          - \mbox{\boldmath p}^2 + m^2
&-& \alpha
          + i \mbox{\boldmath $\beta$} \cdot
                \left( 2 i \mbox{\boldmath A} \cdot
                \mbox{\boldmath p} + \mbox{\boldmath B} \right)
                + \left( 2 i
                        \mbox{\boldmath p} \cdot
                        \mbox{\boldmath A}
                        + \mbox{\boldmath B}
                   \right) \cdot
                {\mbox{\boldmath $\gamma$}}^2 \cdot
                   \left( 2 i
                        \mbox{\boldmath A} \cdot
                        \mbox{\boldmath p}
                        + \mbox{\boldmath B}
                   \right)
\nonumber \\
                &+& 2 i \hspace{0.5mm} \mbox{tr}
                ( {\mbox{\boldmath $\gamma$}}^2 \cdot
                \mbox{\boldmath A} )
        \Big]
= i m^2
        - i \mbox{\boldmath p} \cdot
        \frac{\partial \mbox{\boldmath A}}{\partial s}
        \cdot \mbox{\boldmath p}
        - \frac{\partial \mbox{\boldmath B}}{\partial s}
        \cdot \mbox{\boldmath p}
        - \frac{\partial \mbox{C}}{\partial s}
\eea
and comparing equal powers of $p$ on both sides
we get the following linear differential equations for $A,B$, and $C$,
\bea
\frac{\partial \mbox{\boldmath A}}{\partial s}
&=& \mbox{\boldmath 1}
        + 4 \mbox{\boldmath A} \cdot
        {\mbox{\boldmath $\gamma$}}^2 \cdot
        \mbox{\boldmath A}
\nonumber \\
\frac{\partial \mbox{\boldmath B}}{\partial s}
&=& 2 i \mbox{\boldmath A} \cdot \mbox{\boldmath $\beta$}
        + 4 \mbox{\boldmath A} \cdot
        {\mbox{\boldmath $\gamma$}}^2 \cdot \mbox{\boldmath B}
\nonumber \\
\frac{\partial \mbox{C}}{\partial s}
&=& i \alpha + \mbox{\boldmath $\beta$}
                \cdot \mbox{\boldmath B}
        - i \mbox{\boldmath B} \cdot
                {\mbox{\boldmath $\gamma$}}^2 \cdot
                \mbox{\boldmath B}
        + 2~ \mbox{tr} (
                {\mbox{\boldmath $\gamma$}}^2 \cdot
                \mbox{\boldmath A} )
\eea
The solutions of these equation which satisfy the boundary conditions
(\ref{bc}) are given by
\bea
\mbox{\boldmath A} &=&
\frac{1}{2}\, {\mbox{\boldmath $\gamma$}}^{-1} \cdot
        \tan (2 {\mbox{\boldmath $\gamma$}} s)
\label{eqn15a}
\\
\mbox{\boldmath B}
&=&
- \frac{i}{2}\, {\mbox{\boldmath $\gamma$}}^{-2} \cdot
        \left[ 1 - \mbox{sec} (2 {\mbox{\boldmath $\gamma$}} s) \right]
        \cdot \mbox{\boldmath $\beta$}
\label{eqn15b}
\\
\mbox{C}
&=&
i \alpha s - \frac{1}{2}\, \mbox{tr}
        \left[ \ln \cos (2 {\mbox{\boldmath $\gamma$}} s) \right]
        + \frac{i}{8}\, \mbox{\boldmath $\beta$} \cdot
                {\mbox{\boldmath $\gamma$}}^{-3} \cdot
                \left[ \tan (2 {\mbox{\boldmath $\gamma$}} s)
                - 2 {\mbox{\boldmath $\gamma$}} s \right]
                \cdot \mbox{\boldmath $\beta$}
\label{eqn15c}
\eea
These $\mbox{\boldmath A}, \mbox{\boldmath B}$ and
$\mbox{C}$ determine $G(p)$ when substituted
in (\ref{eqn9}).
The effective Lagrangian is obtained by substituting this
$G(p)$ in (\ref{eqn4}) and carrying out the integration over $m^2$,
\bea
{\cal L}_{eff}
&=& - \frac{i}{2}
        \int_0^{\infty} \frac{ds}{s}
        \int \frac{d^4 p}{(2 \pi)^4}~
        \exp \left\{- i s m^2 + i \mbox{\boldmath p} \cdot
                        \mbox{\boldmath A} \cdot
                        \mbox{\boldmath p}
                        + \mbox{\boldmath B} \cdot
                        \mbox{\boldmath p}
                        + \mbox{C} \right\}
\eea
The Gaussian integral may be evaluated using
\bea
\int d^4 p~
        \exp \left\{i \mbox{\boldmath p} \cdot
                        \mbox{\boldmath A} \cdot
                        \mbox{\boldmath p}
                        + \mbox{\boldmath B} \cdot  \mbox{\boldmath p}
 \right\}
&=&
-\,i \pi^2 \left( \mbox{det} \mbox{\boldmath A} \right)^{-\frac{1}{2}}
        \exp \left[ \frac{i}{4}~ \mbox{\boldmath B} \cdot
                                \mbox{\boldmath A}^{-1} \cdot
                                \mbox{\boldmath B} \right]
\eea
where $\mbox{det} \mbox{\boldmath A}$
is the determinant of the matrix $A^{\mu}_{\nu}$.
Using now (\ref{eqn15a}-\ref{eqn15c}), we have
\bea
{\cal L}_{eff}
&=& - \frac{1}{32 \pi^2}
        \int_0^{\infty} \frac{ds}{s^3}~
        e^{- i s (m^2 - \alpha)}
        \left[
           \mbox{det} \left(
                \frac{2 {\mbox{\boldmath $\gamma$}} s}
                {\sin 2 {\mbox{\boldmath $\gamma$}} s}
                \right)
        \right]^{\frac{1}{2}}
        e^{i l(s)}
\label{eqn1.19}
\eea
where
\bea
l(s)
&=& \frac{1}{4}\, \mbox{\boldmath $\beta$} \cdot
        {\mbox{\boldmath $\gamma$}}^{-3} \cdot
        \left[ \tan ({\mbox{\boldmath $\gamma$}} s)
        - {\mbox{\boldmath $\gamma$}} s \right] \cdot
        \mbox{\boldmath $\beta$}
\eea
The coefficients of the Taylor expansion of the background fields
(\ref{eqn1.6}) determine the effective action on integrating out the
quantum field $\phi$. In particular, an imaginary part of
${\cal L}_{eff}$ may be non-zero depending on the signs of the
eigenvalues of the $\gamma^2$ matrix. When this occurrs, we have
a non-zero probability (\ref{w})
that the external EM fields decay in PS particles.

To the effective Lagrangian in (\ref{eqn1.19}) we should add subtractions
to render it finite at $s=0$. When this is done, we have that
in the limit $\beta \rightarrow 0$,
$\gamma^2 \rightarrow 0$, the effective Lagrangian
${\cal L}_{eff} \rightarrow 0$, as it should be.
In the Appendix A we illustrate the method for the familiar case
of production of charged scalar fields in a constant electric field.

The formulae (\ref{eqn15a}-\ref{eqn15c},\ref{eqn1.19}) differ
by some signs and factors of $i$ from the solutions displayed
in reference \cite{grifols}. There, we presented the formulae for
the case that $f(x)$ had only spatial variation and therefore
$\beta$ and $\gamma^2$ had only $i=1,2,3$ indices. In \cite{grifols} we
used the metric $(+,+,+)$ while in the present paper we
consider spatial as well temporal variation and use the
metric $(+,-,-,-)$. This introduces some changes in intermediate
formulae but of course the final results we get in the present paper
are identical with the final results we got in \cite{grifols}.


\section{Effective EM-fields - PS pair interactions}

In this Section, we show that a coupling of a pseudoscalar to two
photons induces an interaction ${\cal L}_I$ that may lead to PS
production in a background of EM fields.

The generic pseudoscalar-two-photon interaction (see fig.1) can be
written as
\bea
{\cal L}_{\phi\gamma\gamma} = \frac{1}{8} g \phi
        \epsilon^{\mu \nu \rho \sigma} {F}_{\mu \nu}
       {F}_{\rho \sigma}
\label{a2p}
\eea
We should mention that in the
special case where the PS is an axion the coupling $g$ is related
to the mass $m$ of the axion.

We need to evaluate the loop  diagram of the type shown in fig.2 with
infinite number of zero-frequency photon external legs. The imaginary part
of this diagram gives the probability for the decay of the external
electromagnetic field.

To calculate this diagram, we first
evaluate the process $\phi A \rightarrow \phi A$, where $A$ is an external
photon. We use $i {\cal L}_{\phi\gamma\gamma}$ from (\ref{a2p}) in momentum space,
\be
 \frac{1}{4} g \widetilde{\phi}
\epsilon^{\mu \nu \rho \sigma} k_{\mu} \widetilde{A_{\nu}}
        \widetilde{F}_{\rho \sigma}
\ee
The two-photon two-PS interaction is then obtained contracting
the internal photon legs,
\be
 4 \left( \frac{1}{4} g \widetilde{\phi} \right)^2
\epsilon^{\mu \nu \rho \sigma} k_{\mu} \widetilde{F}_{\rho \sigma}
\frac{- i g_{\nu \nu^{\prime}}}{k^2}
\epsilon^{\mu^{\prime} \nu^{\prime} \rho^{\prime} \sigma^{\prime}}
(-k_{\mu^{\prime}})
\widetilde{F}_{\rho^{\prime} \sigma^{\prime}}
\label{p2a0}
\ee
The factor of $4$ in equation (\ref{p2a0}) is
for the four  possible ways of joining the photon legs.
Due to the presence of the $k^2$ term in the denominator, the effective
coupling (\ref{p2a0}) is non-local. However, when we calculate the effective
action for the external EM field the momentum $k$ is integrated over.
One can therefore make use of the identity
\bea
\int d^4 k~ k_{\mu} k_{\mu^{\prime}}\, g( k^2 )
&=& \int d^4 k~
         \frac{ g_{\mu \mu^{\prime}}k^2}{4}\,  g(k^2)
\eea
to simplify (\ref{p2a0}). Thus, we can reduce the effective two PS-two
photon interaction to a local interaction vertex.
Back in configuration space, it is given by
 \bea
{\cal L}_{I}
= - \frac{1}{4} g^2 \phi^2
        {F}_{\mu \nu}
        {F}^{\mu \nu} = \frac{1}{2} g^2 \phi^2
        ( {\bf{E}}^2 - {\bf{B}}^2 )
\label{lif}
\eea
(see fig.3).

With the interaction Lagrangian (\ref{lif}) we can
go back to the formalism of Section II and calculate the
probability density. We can readily identify
$f(x)$ in (\ref{f}),
\be
f(x) =  g^2   ( {\bf{E}}^2 - {\bf{B}}^2 )
\label{EB}
\ee
In order to have a non trivial ${\cal L}_{eff}$, one needs non-zero
second derivatives of the EM fields as they appear in expression
(\ref{EB}). As we said in Section I, depending on the sign
of the corresponding $\gamma^2$ matrix we may have PS production.
We illustrate it in some simple physical situations
in the following sections.


\section{Production of pseudoscalars in dipole magnetic fields}

In a static dipole magnetic field the PS-pair -EM interaction is given
by
\bea
{\cal L}_I
&=&-\frac{1}{2}g^2 B^2 (r) \phi^2
\nonumber \\
&=&
- \frac{1}{2} g^2 \left( B_0^2 z_0^6\, \frac{3 z^2 +r^2}{4 r^8}
\right)
        \phi^2
\eea
where
$B_0$ is the field strength at a point $\vec r_0=(0,0,z_0)$ on the
$z$-axis. We have now
\be
f(\vec r) = - g^2 \left( B_0^2 z_0^6\, \frac{3 z^2 +r^2}{4 r^8} \right)
\ee

Expanding $B^2(r)$ near the point $\vec r_0$
\bea
{\cal L} = \frac{1}{2} \phi \Big[ - \partial^2 - m^2 & + &
  f(\vec r_0) + \left. \frac{\partial f}{\partial x_i}  \right|_{\vec
    r = \vec r_0} (x_i - x_{i0}) \\
 & + & \frac{1}{2}
 \left. \frac{\partial^2 f}{\partial x_i \partial x_j}  \right|_{\vec
    r = \vec r_0} (x_i - x_{i0})(x_j - x_{j0}) \Big] \phi + ...
\eea
we find that the coefficients of the Taylor expansion are given by
\bea
\alpha &=&  f(\vec r_0) =
- {g}^2 B_0^2 \equiv \alpha_m
\label{alpha}
\eea
\bea
(\mbox{\boldmath $\beta$})_i
&=&  \left.
\frac{\partial f}{\partial x_i}  \right|_{\vec
    r = \vec r_0}=
(0,0,6 g^2 B_0^2 z_0^{-1} )
\label{beta}
\eea
and
\bea
(\mbox{\boldmath $\gamma^2$})_{ij}
={1\over 2} \left.{{\partial^2 f\over \partial x_i \partial x_j }}
\right|_{\vec r = \vec r_0} =
\frac{3 g^2 B_0^2}{4 z_0^2}
\pmatrix{
5 & 0 & 0 \cr
0 &5 & 0 \cr
0 & 0 & -28
}
\equiv
\pmatrix{
a_m^2 & 0 & 0 \cr
0 & a_m^2 & 0 \cr
0 & 0 & - b_m^2
}
\label{g2_m}
\eea

Therefore we find using the notation and formalism of Section II that
 the effective  action on integrating out the PS field is
given by
\be
{\cal L}_{eff}
= - \frac{1}{32 \pi^2}
        \int_0^{\infty} \frac{ds}{s^3}~
        e^{- i s (m^2 - \alpha_m)}
\frac{2 a_m s}{\sinh 2 a_m s}
\sqrt{ \frac{2 b_m s}{\sin 2 b_m s} }
        e^{i l_m(s)}
\label{ldipole}
\ee
with
\be
l_m(s) = \lambda_m (b_m s - \tan b_m s)
\ee
\be
 \lambda_m  = \frac{9 g^4 B_0^4}{z_0^2} \frac{1}{b_m^3} =
 \frac{3}{7  \sqrt{ 21} } g B_0 z_0
\ee

The imaginary part of the expression (\ref{ldipole}) can be
performed by enclosing the simple poles at
$s= - i n \pi (2 a_m)^{-1}$, $n=0,1,...$,
with a contour from
below. We get
\be
\mbox{Im}~ {\cal L}_{eff} =
\frac{1}{8\pi^\frac{5}{2}}\,
a_m^\frac{3}{2}\, b_m^\frac{1}{2}
 \sum_{n=1}^{\infty}
        (-1)^{n + 1} C^{(m)}_n e^{ - n \pi/2 \eta_m}
\ee
\be
C^{(m)}_n =
n^{-\frac{3}{2}}
\left[ \sinh n   \frac{b_m}{a_m} \pi\right]^{-\frac{1}{2}}\,
e^{ \tilde l_m }
\label{c_m}
\ee
with
\be
\eta_m = \frac{ a_m}{m^2-\alpha_m } =
\frac{\sqrt{15}}{2} \left( \frac{m^2 z_0}{g B_0} +
z_0 g B_0 \right)^{-1}
\ee
and
\be
\tilde l_m = \lambda_m \left[ n \frac{b_m}{a_m} \frac{ \pi}{2} -
\tanh  n \frac{b_m}{a_m} \frac{ \pi}{2} \right] \label{l_m} \ee In
(\ref{c_m}) and (\ref{l_m}) we can put $b_m/a_m=\sqrt{28/5}$. The
main contribution to the above integral comes from the $n=1$ term
and we find that $w$ is given by the expression
\bea
\label{w_m} 
w&=& 0.036
{g^2 B_0^2  \over
z_0^2} 
e^{2.72  \lambda_m} \
e^{ - \pi/ 2 \eta_m }
 \eea
 
The probability $w$ of field decay is extremely suppressed for realistic 
parameters. To illustrate this,
let us choose a mass $m$ and a coupling $g$ consistent with
the axion window: 
 \bea
\label{window}
 m&\sim& 10^{-3}\  {\rm eV}
 \nonumber\\
 g &\sim& 10^{-13}\ {\rm GeV}^{-1}
 \eea
Also, let us choose 
 \bea
\label{field}
 B_0&=& 1\ {\rm Tesla} \nonumber\\
 z_0 &=& 10\ {\rm cm}
 \eea
With these values, we get
 \bea
 \eta_m &\sim& 10^{-19}
 \eea
and since $\eta_m$ appears in the exponential in (\ref{w_m}), 
we see that the probability $w$ is indeed extremely suppressed. 
For a dipole magnetic
field to be unstable and decay into
axions, one needs $\eta_m \gtrsim 1$, but this
would correspond either to unrealistic values for the external field
parameters (\ref{field}) or to excluded values for the axion
mass and coupling (\ref{window}). For non-axion models, $g$ and 
$m$ are not related (still there are restrictions on these
parameters, see ref.\cite{mt}). One could have $\eta_m \sim 1$ 
by tuning $g$ and $m$. Imposing that the field (\ref{field}) does
not decay into pseudoscalars leads to the constraint
 \bea
 \left( \frac{m}{ 10^{-12}\, {\rm eV}} \right)^2
 &\gtrsim& 
 \frac{g}{ 10^{-13}\, {\rm GeV}^{-1} }
 \eea

\section{Production of pseudoscalars in a cylindrical capacitor}

The modulus of the electric field inside a cylindrical capacitor whose
axis lies along
the $z$-axis depends only on $\rho=(x^2+y^2)^\frac{1}{2}$,
\be
E(\rho)= \frac{\lambda}{2 \pi} \frac{1}{\rho}
\ee
with $\lambda$ the linear electric charge density.

The bilinear interaction term (\ref{lif}) is
\bea
{\cal L}_I
&=& \frac{1}{2} g^2 E^2 (\rho) \phi^2(x)
\nonumber \\
&=&
\frac{1}{2} g_c^2  \left( \frac{1}{\rho^2} \right) \phi^2(x)
\eea
where
$g_c \equiv \lambda g/2 \pi$. The corresponding function $f(x)$
is
\be
f(\rho)= g_c^2 \left( \frac{1}{\rho^2} \right)
\ee
Expanding the fields near some reference point $(x_0,y_0,z_0)$
with $\rho_0=(x_0^2+y_0^2)^\frac{1}{2}$
\bea
{\cal L}
&=& \frac{1}{2}
        \phi \left[
                - \partial^2 - m^2
                + f(\rho_0)
                + \left. \frac{\partial f}{\partial \rho}
                        \right|_{\rho = \rho_0} (\rho - \rho_0)
                + \left. \frac{1}{2} \frac{\partial^2 f}{\partial \rho^2}
                        \right|_{\rho = \rho_0} (\rho - \rho_0)^2
             \right] \phi + ...
\eea

It can be written as in (\ref{eqn1.6}) with
\bea
\alpha &=&  f(\rho_0) \equiv \alpha_c
\eea
\bea
(\mbox{\boldmath $\beta$})_i
&=& \frac{f^{\prime}}{\rho_0}
        (x_0, y_0)
\eea
and
\bea
(\mbox{\boldmath $\gamma^2$})_{ij}
&=& \frac{f^{\prime}}{2 \rho_0^3}
\pmatrix{
y_0^2  & - x_0 y_0  \cr
- x_0 y_0 & x_0^2
}
+
\frac{f^{\prime \prime}}{2 \rho_0^2}
\pmatrix{
x_0^2 & x_0 y_0  \cr
x_0 y_0 & y_0^2
}
\label{eqng2c}
\eea
where primes denote derivatives with respect to $\rho$
taken at $\rho=\rho_0$. In the above formulae, the spatial indices
run over $1,2$.

Introducing the explicit form of $f$, we get
\bea
\alpha_c &=&
 \frac{g_c^2}{\rho_0^2}
\eea
\bea
(\mbox{\boldmath $\beta$})_i
&=& - \frac{2 g_c^2}{\rho_0^4}
        (x_0, y_0)
\eea
and the $\gamma^2$ matrix (\ref{eqng2c}) reads
\bea
(\mbox{\boldmath $\gamma^2$})_{ij}
&=& \frac{g_c^2}{\rho_0^6}
\pmatrix{
-y_0^2 + 3 x_0^2 & 4 x_0 y_0  \cr
4 x_0 y_0 & -x_0^2  + 3 y_0^2
}
\eea

Next, we diagonalise $\gamma^2$
by rotating the coordinates with an orthogonal matrix. For example,
we can use
\bea
(\mbox{\boldmath O})_{i j}
= \frac{1}{\rho_0}
\pmatrix{
 x_0 & y_0  \cr
-y_0 & x_0
}
\eea

In diagonal form we have
\bea
({\mbox{\boldmath $\gamma$}}^2_D)_{i j}
= \frac{g_c^2}{\rho_0^4}
\pmatrix{
3 & 0  \cr
0 & -1
}
\equiv
\pmatrix{
  a_c^2  & 0 \cr
0 & - b_c^2
}
\eea
We need $\vec{\beta}$ in the diagonal basis given by
$\vec{\beta_D} = O \cdot \vec{\beta}$. We get
\bea
(\mbox{\boldmath $\beta$}_D)_i
&=& \left( - 2 g_c^2 \rho_0^{-3}, 0 \right)
\eea

The expression for ${\cal L}_{eff}$ (\ref{eqn1.19}) finally reads
\bea
{\cal L}_{eff}
&=& - \frac{1}{32 \pi^2}
        \int_0^{\infty} \frac{ds}{s^3}~
        e^{- i s (m^2 - \alpha_c)}
        \sqrt{
            \frac{2 a_c s}{\sinh 2 a_c s}
        }~
\sqrt{
        \frac{2 b_c s}{\sin 2 b_c s}
        }~
       e^{i l_c(s)}
\label{leef_c}
\eea
where
\bea
l_c(s)
&=&
    \lambda_c   \left( a_c s - \tanh a_c s \right) \\
\lambda_c &=& g_c^4 \rho_0^{-6} a_c^{-3} =
\frac{g_c}{3 \, \sqrt{3}}
\eea

We are not able to perform the integration in (\ref{leef_c})
by the procedure of extending $s$ to the complex plane, as we
have done in section IV. The reason is the presence of
essential singularities contained in $l_c(s)$. 
(For a discussion
of the implications of essential singularities in the context
of QED pair production calculations at finite temperature,
see ref.(\cite{dittrich}.)

We shall calculate  the integral (\ref{leef_c}) numerically.
We make the change of variables
\bea
x
&=&
(m^2-\alpha_c - \lambda_c a_c )\, s
\eea
so that
\bea
w = 2\, \mbox{Im}~ {\cal L}_{eff}
&=&
\frac{1}{16 \pi^2}\frac{1}{(m^2-\alpha_c - \lambda_c a_c )^2 }\ I_c
\eea
where
\bea\label{numc}
I_c
&=&
 \int_0^{\infty} \frac{dx}{x^3}~\sin(\phi_c)~
\left(\sqrt{
            \frac{2 \eta_c x}{\sinh 2  \eta_c x}
        }~
\sqrt{
        \frac{2  \eta_c x/\sqrt 3}{\sin  (\eta_c x/ \sqrt 3) }
        } + \frac{2}{9} x^2 - 1 
\right)
\eea
We have introduced the necessary subtractions and defined
\bea
\eta_c
&=&
\frac{a_c}{m^2-\alpha_c - \lambda_c a_c}
\eea
and
\bea
\phi_c
&=&
x + \lambda_c \tanh \eta_c x
\eea

As it stands, $I_c$ depends on the two adimensional parameters
$\lambda_c$ and $\eta_c$, which reflect its dependence on both the
strength of the interaction and on the mass of the scalar particles. In
order to explore numerically this two-dimensional space we should
focus on those regions for which the results make physical sense. This
is even more so because a blind computation of the integral leads
easily to wild fluctuations due to the violent oscillations of the
integrand for large domains of parameter space.

Since the electric field should be inhomogeneous on scales of the order
of a particle's Compton wavelength, we have
\bea
\label{tg1}
\left| E^{-1} \left( {dE \over d\rho } \right)_0 m^{-1}\right| \  > \ 1
\eea
In our case, this means
\bea
\label{tg2}
\rho _{0} m <1 
\eea
Moreover, we should restrict our survey to subcritical conditions, i.e.
conditions such that we find ourselves below the point where
catastrophical pair-production starts in the earnest and vacuum
breakdown occurs. The system will be subcritical whenever the field
energy stored in a volume of size $m^{-3}$, 
and which is converted into a PS pair,  
is at most of the order of twice the scalar particle rest
mass. A crude back of an envelope estimate gives,
\bea
\label{tg3}
\lambda_c^2 /m^{2}\rho_{0}^{2}\lesssim {\cal O }(1)
\eea

Using both restrictions as a rough guide, we perform the numerical
integration of $I_c$ as a function of $\eta_c ^{-1}$ for 
$\eta_c$ and $\lambda_c$ in the ballpark of the values required by
(\ref{tg2}) and (\ref{tg3}).
The results for two values of $\lambda_c$ are displayed in 
figs.4 and 5. 
The curves are
accurately fit by an analytic expression of the form,  
\bea
\label{tg4}
I_c=A\eta_c ^2 e^{-k/\eta_c}
\eea
with $A$ and $k$ depending on  $\lambda_c$. ($A$ and $k$ are
positive.)
Equation (\ref{tg4}) has a form that closely resembles the classical 
Schwinger result
and characterizes a typical non-perturbative process.

The electric field would break down into pseudoscalars if the exponent
in (\ref{tg4}) becomes small. Again we choose the values consistent
with the axion window (\ref{window}) and for the field parameters
we take
\bea
E_0 &\sim& 10^4\ {\rm V/m}
\nonumber\\
\rho_0 &\sim& 0.1\ {\rm cm}
\label{ccap}
\eea
This leads to  
\bea
\eta_c&\sim& 10^{-22}
\eea 
which implies a large negative value of the exponent in  (\ref{tg4})
that suppresses field decay. 

\section{Production of pseudoscalars in a spherical capacitor}

The modulus of the electric field inside a spherical capacitor depends only
on $r=|\vec r\, |$,
\be
E(r)= \frac{Q}{4 \pi} \frac{1}{r^2}
\ee
where $Q$ is the electric charge.

The bilinear interaction term is then
\bea
{\cal L}_I
&=& \frac{1}{2} g^2 E^2 (r) \phi^2(x)
\nonumber \\
&=&
\frac{1}{2} g_s^2 \left( \frac{1}{r^4} \right) \phi^2(x)
\eea
where
$g_s = Q g/4 \pi$.

The corresponding function $f(x)$ depends only on $r$,
\be
f(r)=g_s^2 \left( \frac{1}{r^4} \right)
\ee
Expanding the fields near some reference point with $(x_0,y_0,z_0)$
with modulus $r_0$
\bea
{\cal L}
&=& \frac{1}{2}
        \phi \left[
                - \partial^2 - m^2
                + f(r_0)
                + \left. \frac{\partial f}{\partial r}
                        \right|_{r = r_0} (r - r_0)
                + \left. \frac{1}{2} \frac{\partial^2 f}{\partial r^2}
                        \right|_{r = r_0} (r - r_0)^2
             \right] \phi + ...
\eea

In Cartesian coordinates it can be written as in (\ref{eqn1.6})
with
\bea
\alpha &=&  f(r_0) \equiv \alpha_s
\eea
\bea
(\mbox{\boldmath $\beta$})_i
&=& \frac{f^{\prime}}{r_0}
        (x_0, y_0, z_0)
\eea
and
\bea
(\mbox{\boldmath $\gamma^2$})_{ij}
&=& \frac{f^{\prime}}{2 r_0^3}
\pmatrix{
y_0^2 + z_0^2 & - x_0 y_0 & - x_0 z_0 \cr
- x_0 y_0 & x_0^2 + z_0^2 & - y_0 z_0 \cr
- x_0 z_0 & - y_0 z_0 & x_0^2 + y_0^2
}
+
\frac{f^{\prime \prime}}{2 r_0^2}
\pmatrix{
x_0^2 & x_0 y_0 & x_0 z_0 \cr
x_0 y_0 & y_0^2 & y_0 z_0 \cr
x_0 z_0 & y_0 z_0 & z_0^2
}
\eea
where primes denote derivatives with respect to $r$
taken at $r=r_0$. Introducing the form of $f(x)$, we get
\bea
\alpha_s &=&
 \left( \frac{g_s^2 }{r_0^4} \right)
\eea
and
\bea
(\mbox{\boldmath $\beta$})_i
&=& - \frac{4 g_s^2}{r_0^6}
        (x_0, y_0, z_0)
\eea
and
\bea
(\mbox{\boldmath $\gamma^2$})_{ij}
&=& \frac{2 g_s^2}{r_0^8}
\pmatrix{
5 x_0^2 - y_0^2 - z_0^2 & 6 x_0 y_0 & 6 x_0 z_0 \cr
6 x_0 y_0 & - x_0^2 + 5 y_0^2 - z_0^2 & 6 y_0 z_0 \cr
6 x_0 z_0 & 6 y_0 z_0 & - x_0^2 - y_0^2 + 5 z_0^2
}
\label{eqn3.3b}
\eea

The $\gamma^2$ matrix (\ref{eqn3.3b}) can be diagonalised
by rotating the coordinates with an orthogonal matrix. For example,
we can use
\bea
(\mbox{\boldmath O})_{i j}
= \frac{1}{r_0 d_0}
\pmatrix{
- z_0 r_0 & 0 & x_0 r_0 \cr
- x_0 y_0 &
        d_0^2 & - y_0 z_0 \cr
x_0 d_0 & y_0 d_0 & z_0 d_0
}
\eea
where $d_0 = \sqrt{ x_0^2 + z_0^2 }$.

In diagonal form we have
\bea
({\mbox{\boldmath $\gamma$}}^2_D)_{i j}
= \frac{2 g_s^2}{r_0^6}
\pmatrix{
-1 & 0 & 0 \cr
0 & -1 & 0 \cr
0 & 0 & 5
}
\equiv
\pmatrix{
 - b_s^2  & 0 & 0 \cr
0 & - b_s^2 & 0\cr
0 & 0 & a_s^2
}
\eea
One must also use $\vec{\beta}$ in the diagonal basis given by
$\vec{\beta_D} = O \cdot \vec{\beta}$. We get
\bea
(\mbox{\boldmath $\beta$}_D)_i
&=& \left( 0, 0, - 4 g_c^2 r_0^{-5} \right)
\eea
The expression for ${\cal L}_{eff}$ (\ref{eqn1.19}) is given in
this case by
\bea
{\cal L}_{eff}
&=& - \frac{1}{32 \pi^2}
        \int_0^{\infty} \frac{ds}{s^3}~
        e^{- i s (m^2 - \alpha_s)}
        \sqrt{
            \frac{2 a_s s}{\sinh 2 a_s s}
        }~
        \frac{2 b_s s}{\sin 2  b_s s}
        e^{i l_s(s)}
\label{eqn3.7a}
\eea
where
\bea
l_s(s)
&=&
\lambda_s    \left( a_s s - \tanh a_s s \right) \\
\lambda_s &=& \frac{4 g_s^4}{r_0^{10}}\frac{1}{a_s^3}
=  \frac{2 }{5 \, \sqrt{10}}\,  \frac{g_s}{r_0}
\eea

As in the precedent section, we cannot perform the integration in 
(\ref{eqn3.7a}) by extending $s$ to the complex plane since 
again there are essential singularities and we do a numerical
integration. The procedure is very similar. It is convenient 
to change variables,
\bea
x
&=&
(m^2-\alpha_s - \lambda_s a_s )\, s
\eea
so that
\bea
w = 2\, \mbox{Im}~ {\cal L}_{eff}
&=&
\frac{1}{16 \pi^2}\frac{1}{(m^2-\alpha_s - \lambda_s a_s )^2 }\ I_s
\eea
where
\bea\label{nums}
I_s
&=&
 \int_0^{\infty} \frac{dx}{x^3}~\sin(\phi_s)~
\left(\sqrt{
            \frac{2 \eta_s x}{\sinh 2  \eta_s x}
        }~
        \frac{2  \eta_s x/\sqrt 5}{\sin  (\eta_s x/ \sqrt 5) }
         + \frac{1}{5} x^2 - 1 
\right)
\eea
We have introduced the necessary substractions and defined
\bea
\eta_s
&=&
\frac{a_s}{m^2-\alpha_s - \lambda_s a_s}
\eea
and
\bea
\phi_s
&=&
x + \lambda_s \tanh \eta_s x
\eea

Here we follow a similar strategy as before to pinpoint the relevant
parameter space. We get similar restrictions:
\bea
\label{tg5}
r_0 m<1 \ \ \ \ {\rm and} \ \ \ \  
\lambda_s ^2/r_0^{2}m^2\lesssim {\cal O} (1)
\eea
In figs.6 and 7 we display  $I_s$ as a function of $\eta_s^{-1}$, 
for a couple of values of $\lambda_s$. Again, we see that the 
approximate behaviour
is that of a decreasing exponential,
 \bea
\label{tg6}
I_s
&=&
A \eta_s^2 e^{-k/\eta_s}
\eea
with positive constants $A$ and $k$ (that depend on $\lambda_s$).

When one considers 
realistic values for the field and PS parameters it turns out that the 
probability of field breakdown is extremely suppressed as it happens in
the cases that have been analyzed in the precedent sections,
namely the dipole magnetic field and the cylindrical capacitor.


\section{Conclusions and Final Remarks}

In the presence of strong external fields, the physical vacuum breaks
down because particle-antiparticle pairs are being pumped out of it
at the expense of field energy. The case of a strong uniform electric
field spontaneously creating electron-positron pairs is the best
known (QED) example for this phenomenon. Such process is of a
non-perturbative nature and the QED case has been solved exactly by
Schwinger and others \cite{schwinger}.
Their solution however does not include the
backreaction on the external field exerted by the presence of the
produced $e^+e^-$ pairs. Clearly, creation of pairs requires the
supply of mass energy and kinetic energy which must be furnished by
the external field. A balanced energy budget is therefore only
possible through a corresponding reduction of the energy stored in the
field. Because electrons and positrons carry charge they will fly to
the external sources of the field and thus the field (and hence its
energy) will diminish. So, unless from the outside the field is
restored, the pair production process cannot be indefinitely
sustained. If nothing is done from the outside a catastrophic
breakdown of the initially strong (critical) electric field will
inevitably follow.

In the present paper we dealt with pseudoscalar particles.
Pseudoscalars are fundamental ingredients of many completions of
Particle Physics models. Examples run from axions to superlight
partners of gravitinos. In the previous sections we have derived the
probability for pair production of PS in electric and magnetic
fields. Contrary to the QED case mentioned above, constant fields do
not cause the disruption of the vacuum. Field gradients are
necessary for the phenomenon to occur. Hence, we studied PS pair
production in inhomogeneous fields. We have calculated the
probability in a general case and based our computation on an
effective action formalism formulated by Brown and Duff. We then
have applied the general formulae to a few specific cases: PS
production  in a magnetic dipole field and
between the plates of a charged capacitor (either
cylindrical or spherical). Again,
backreaction was ignored and therefore adequate boundary
conditions were implicitly assumed that take into account the fact
that pairwise creation of PS requires field energy to be depleted.

In the three cases studied, we found that our probability shows the
non-perturbative behaviour 
$w \sim exp(- const\,\times \, m^2 /g)$ expected for subcritical
fields.
Finally, we should point out
that in a previous paper \cite{grifols} we erroneously estimated
axion emission in
the Coulomb field of an atomic nucleus. This result is incorrect
because we overlooked the question of appropriate boundary conditions
that guarantee energy conservation and which are clearly not met in
this microscopic system.

\begin{acknowledgments}
We would like to thank G. Raffelt for pointing out a conceptual
inconsistency in a previous version of this work and for the
illuminating discussions that followed. We also thank the
referees for their valuable comments, especially for raising the 
question concerning essential
singularities and for drawing our attention to reference \cite{dittrich}.
Two of us (J.A.G. and
E.M.) have partial support from the CICYT Research Project
AEN99-0766.
\end{acknowledgments}

\appendix

\section{
Decay of a constant electric field into charged scalars
}

We start with the equation for the Green's function $G(p)$
\bea
\left[ - (p - eA)^2 + m^2 \right] G(p)
&=& 1
\nonumber \\
\left[ - p^2 + m^2
        + e ( A^{\mu} p_{\mu} + p_{\mu} A^{\mu} )
        - e^2 A^2 \right] G(p)
&=& 1
\label{a1}
\eea
We assume constant ${\bf E}$ and ${\bf B}$ fields.
The vector potential can be choosen as
\bea
A_{\mu}
&=& - \frac{1}{2} F_{\mu \nu} x^{\nu}
\rightarrow
\frac{i}{2} F_{\mu \nu}
        \frac{\partial}{\partial p_{\nu}}
\eea
When inserted in (\ref{a1}), one gets an equation of the form
given in (\ref{eqn8}), except for a term
\be
{F^\mu}_\nu p_\mu  \frac{\partial}{\partial p_{\nu}} G(p)
\ee
that leads to an expression containing
\be
{F^\mu}_\nu p_\mu p_\alpha A^{\alpha \mu}
\ee
The antisymmetry of $F$ and the fact that $F$ and $A$ commute
makes this term vanish. Then our equation is
\bea
\left[ - p^2 + m^2
        + \frac{e^2}{4} F_{\mu \nu} F^{\mu}_{\rho}
        \frac{\partial^2}{\partial p_{\nu} \partial p_{\rho}}
        \right]
        G(p)
&=& 1
\eea
and with our definitions in (\ref{eqn1.6}), $\beta = 0$
and
\bea
\gamma^2_{\nu \rho}
&=&
- \frac{e^2}{4}
        F_{\mu \nu} F^{\mu}_{\rho}
\nonumber \\
&\equiv&
- \frac{e^2}{4} F^2_{\nu \rho}
\eea
Let us work out the special case of a constant ${\bf E}$ field.
We have
\bea
(\gamma^2)_{\nu \rho}
&=&
- \frac{e^2}{4}
\pmatrix{
E^2 & 0 & 0 & 0 \cr
0 & 0 & 0 & 0 \cr
0 & 0 & 0 & 0 \cr
0 & 0 & 0 & - E^2
}
\eea
(We have chosen the $z$-direction as the direction of ${\bf E}$).
The eigenvalues of ${\gamma^2}^{\nu}_{\rho}$ are negative so
\bea
\left[ \mbox{det} \left(
                \frac{2 {\mbox{\boldmath $\gamma$}} s}
                {\sin 2 {\mbox{\boldmath $\gamma$}} s}
                \right)
        \right]^{\frac{1}{2}}
&=&
\frac{e E s}{\sinh e E s}
\eea
and
\bea
{\cal L}_{eff}
&=& - \frac{1}{32 \pi^2}
        \int_0^{\infty} \frac{ds}{s^3}~
        e^{- i s m^2}
        \frac{e E s}{\sinh e E s}
\eea
        To this expression for ${\cal L}_{eff}$ one should add a
subtraction to make it finite at $s=0$.
When this is done, ${\cal L}_{eff} \rightarrow 0$
when $eE \rightarrow 0$.

The probability of scalar production can now be calculated
using (\ref{eqn1.5}).
The integral can be calculated by contour integration by closing the
real axis with a contour on the negative imaginary plane. This contour
encloses poles at $s= -i n \pi/e E$ which contribute to the integral.
The final result for the constant electric field decay probability density
is
\bea
w=
 \frac{\alpha E^2}{2 \pi^2}
        \sum_{n=1}^{\infty}
        \frac{(-1)^{n + 1}}{n^2}
        \exp \left( - \frac{n \pi m^2}{e E} \right)
\eea
which coincides with the well-known formula found in textbooks
\cite{iz}.


\newpage

\begin{figure}
 \label{fig1}
\centering
\includegraphics[width=5cm]{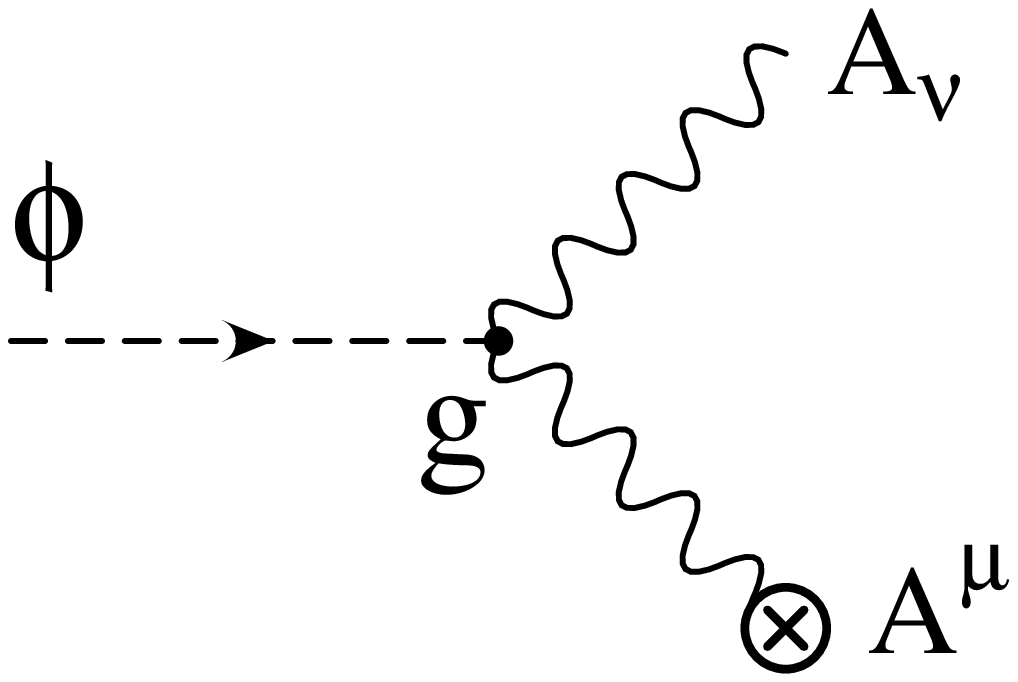}
\caption{PS-two-photon interaction.}
\end{figure}

\begin{figure}
 \label{fig2}
\begin{center}
\includegraphics[width=7cm]{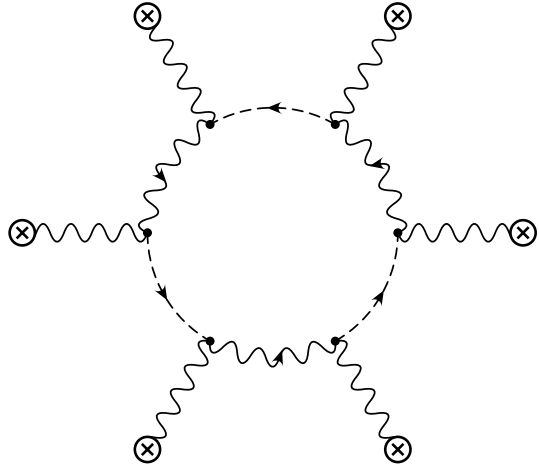}
\end{center}
\caption{Loop diagram showing infinite number of photon
external legs.}
\end{figure}

\begin{figure}
 \label{fig3}
\begin{center}
\includegraphics[width=5cm]{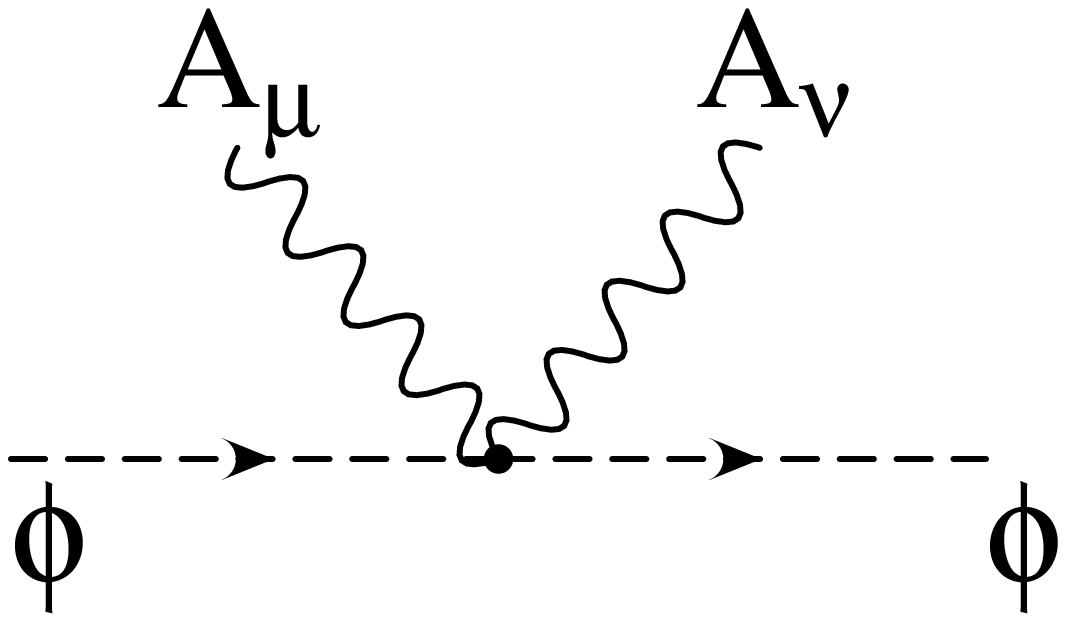}
\end{center}
\caption{Two PS-two photon interaction.}
\end{figure}

\begin{figure}
 \label{fig4}
\begin{center}
        \includegraphics[width=8cm]{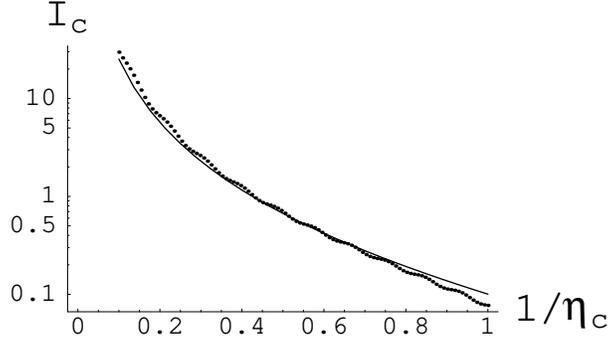}
\end{center}
\caption{$I_c$ as a function of $1/\eta_c$ 
for the value $\lambda_c=0.1$. Dotted line: $I_c$ obtained
by numerical integration of (\ref{numc}). 
Full line: $I_c$ as given by
(\ref{tg4}) with $A=0.28$ and $k=1.026$}
\end{figure}

\begin{figure}
 \label{fig5}
\begin{center}
        \includegraphics[width=8cm]{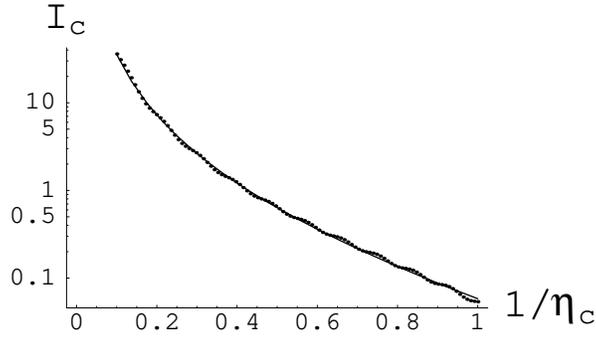}
\end{center}
\caption{Same as  fig.4 for
$\lambda_c=0.4$ and $A=0.43$ and $k=2.1$}
\end{figure}

\begin{figure}
 \label{fig6}
\begin{center}
        \includegraphics[width=8cm]{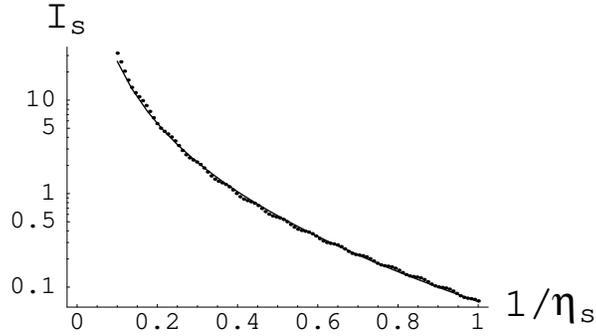}
\end{center}
\caption{$I_s$ as a function of $1/\eta_s$ 
for the value $\lambda_s=0.2$. Dotted line: $I_s$ obtained 
by numerical integration of (\ref{nums}). 
Full line: $I_s$ given by
(\ref{tg6}) with $A=0.30$ and $k=1.46$}
\end{figure}

\begin{figure}
 \label{fig7}
\begin{center}
        \includegraphics[width=8cm]{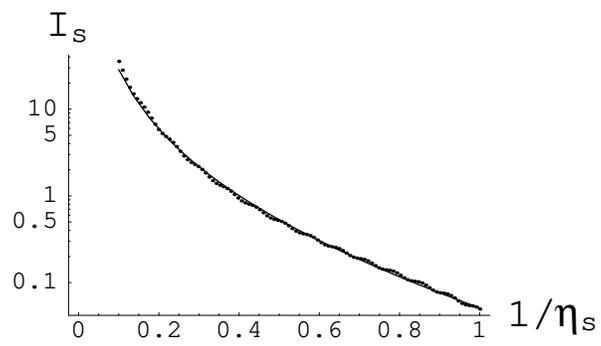}
\end{center}
\caption{Same as fig.6  for
$\lambda_s=0.5$ and 
$A=0.35$ and $k=1.92$}
\end{figure}

\end{document}